%% file: main.tex
\documentclass[reprint,
superscriptaddress,
preprintnumbers,
nofootinbib,
amsmath,amssymb,
prd]{revtex4-1}

\usepackage{graphicx}
\usepackage{dcolumn}
\usepackage{bm}
\usepackage{hyperref}
\usepackage{textgreek}
\usepackage{epsfig,amsmath,amssymb,verbatim,mathrsfs,array,layout,textcomp,amssymb,latexsym,slashed,graphicx,color,mathtools,tikz}
\usepackage{siunitx}
\usepackage{multirow}

\begin{document}

\preprint{FERMILAB-PUB-20-088-AE-E}

\title{Constraints on low-mass, relic dark matter candidates from a surface-operated SuperCDMS single-charge sensitive detector}

\input{authors}

\date{\today}


\begin{abstract}
This article presents an analysis and the resulting limits on light dark matter inelastically scattering off of electrons, and on dark photon and axion-like particle absorption, using a second-generation SuperCDMS high-voltage eV-resolution detector. The 0.93~gram
Si detector
achieved a 3\,eV phonon energy resolution;
for a detector bias of 100\,V, this corresponds to a charge resolution of 3\,\% of a single electron-hole pair.
The energy spectrum is reported from a blind analysis with 1.2~gram-days of exposure acquired in an above-ground laboratory.
With charge carrier trapping and impact ionization effects incorporated into the dark matter signal models,
the dark matter-electron cross section $\bar{\sigma}_{e}$ is constrained for dark matter masses from 0.5--$10^{4}$\,MeV$/c^{2}$; in the mass range from 1.2--50\,eV$/c^{2}$ the dark photon kinetic mixing parameter $\varepsilon$ and the axioelectric coupling constant $g_{ae}$ are constrained. The minimum 90\,\% confidence-level upper limits within the above mentioned mass ranges are $\bar{\sigma}_{e}\,=\,8.7\times10^{-34}$\,cm$^{2}$, $\varepsilon\,=\,3.3\times10^{-14}$, and $g_{ae}\,=\,1.0\times10^{-9}$. 


\end{abstract}

\maketitle

\section{Introduction}

During the past two decades, many significant constraints on weakly interacting massive particle (WIMP) dark matter (DM) for masses above $10$\,GeV$/c^{2}$ have been published (e.g.~\cite{Aprile:2018,Akerib:2017,Agnes:2018,Agnese:2019,Ajaj:2019,Armengaud:2019,Abdelhameed:2019,Cui:2017,PhysRevD.94.082006,PhysRevD.100.022001} and references therein). In contrast to the standard WIMP, well-motivated alternative models at masses below a few GeV$/c^{2}$ that require at least one new gauge boson to satisfy the observed relic density remain relatively unexplored~\cite{Battaglieri:2017}. We undertook a search for such candidates with a SuperCDMS high-voltage eV-resolution (HVeV) detector~\cite{Agnese:2018,Romani:2018,Hong:2019}.  We constrain three DM candidates: (1)
light DM $\chi$, referring to thermal DM particles that inelastically scatter with electrons via a new dark sector force mediator~\cite{Boehm:2004,Pospelov:2008b}; (2)
dark photons $V$ that kinetically mix with Standard Model (SM) photons~\cite{Okun:1982,Holdom:1986,Galison:1984}; and (3)
axion-like particles (ALPs) that are
absorbed by an atom via the axioelectric effect~\cite{Peccei:2008,Svrcek:2006,Derevianko:2010}. These candidates can create electron-hole ($e^{-}h^{+}$) pairs in the phonon-mediated cryogenic silicon HVeV detector. 


In a prior work~\cite{Agnese:2018}, we undertook an above-ground search with a first-generation Si HVeV detector. Contemporaneously, the SENSEI Collaboration reported an underground search with Skipper CCDs~\cite{Abramoff:2019}.
Both works excluded new parameter space for light DM scattering and dark photon absorption in similar mass ranges, but did not report on the axioelectric coupling, which is most strongly constrained by astronomical observations~\cite{An:2015,Viaux:2013,Tanabashi:2018,Bertolami:2014}.
In this work, we analyze a slightly larger above-ground exposure of 1.2~gram-days of a second-generation Si HVeV detector with the same dimensions but modified sensor design compared to~\cite{Agnese:2018}, leading to an improved phonon energy resolution as good as $\sigma_{E}=3$\,eV at the single-$e^{-}h^{+}$-pair energy (3\,\% charge resolution for a 100\,V bias).
Using signal models that include the contributions from charge carrier trapping and impact ionization, we report the constraints obtained from a blind analysis on $\chi$ scattering for DM masses from 0.5--$10^{4}$\,MeV$/c^{2}$, as well as $V$ and ALP absorption for masses from 1.2--50\,eV$/c^{2}$.

\section{Experimental setup}

The data were acquired in a surface laboratory at Northwestern University (Evanston, IL), with the overburden and environmental radioactivity of a typical steel-concrete building. The detector is made of a 0.93~gram high-purity Si crystal ($1\times1\times0.4$\,cm$^{3}$). We clamped the detector between two printed circuit boards for thermal sinking and electrical connections. To reject correlated environmental noise, we installed an anti-coincidence (veto) detector adjacent to the HVeV detector
in the same light-tight copper housing mounted to the cold finger of an Adiabatic Demagnetization Refrigerator (ADR).
More information about the detector setup and the infrared radiation shield is available in Ref.~\cite{Hong:2019}.

SuperCDMS HVeV detectors measure phonons created by particle interactions in the Si crystals with two distributed channels of Quasiparticle-trap-assisted Electrothermal-feedback TESs\footnote{Transition edge sensors \cite{Irwin:2005}.} (QETs) \cite{Hong:2019}.
The QETs fabricated on this device have a superconducting transition temperature $T_{c}\approx65$\,mK.
One QET channel is a square with a sensitive area of 0.5\,cm$^{2}$, and the other is a surrounding frame of equal area.
Both are
on the detector's top surface.
On the bottom surface, an aluminum grid with 5\,\% surface coverage provides a uniform electric field between the two surfaces. The veto detector consists of a single TES on a thin Si wafer that is identical to the TES described in Ref.~\cite{fink2020characterizing} but with $T_c\approx52$\,mK.

We cycled the ADR daily from 4\,K to the base temperature and then regulated it at 50--52\,mK during data taking to obtain a 10--12~hour/day hold time~\cite{Hong:2019}.
To induce Neganov-Trofimov-Luke (NTL) amplification~\cite{Neganov:1985,Luke:1988},
the aluminum grid was biased at $V_{\textrm{bias}} = 100$\,V
while the QETs and detector housing
were held at
ground potential.
At the start of each daily cycle, we set the operating point of each QET to $\sim300~\textrm{m}\Omega$ (about 45\,\% of its normal-state resistance) and recalibrated the detector using a 635-nm laser that was fiber-coupled from room temperature.
Each QET was read out with a DC superconducting quantum interference device (SQUID) at 1\,K operated in a flux-locked feedback loop, and the signals were digitized continuously at 1.51\,MHz.
The laser intensity was adjusted to achieve a mean number of $e^{-}h^{+}$ pairs per pulse between 1 and 4, which produced enough events for calibration up to seven $e^{-}h^{+}$ pairs per event.
We also took a dedicated laser dataset in which we varied the crystal temperature but held the QETs at their nominal operating point; we used this dataset to reconstruct the temperature dependence of the QET responsivity.



\section{Data Collection and Event Reconstruction}

A raw exposure of 3.0~gram-days was collected over 7~days during April--May of 2019.
By partitioning the continuous-acquisition data stream  into 10-second long intervals,
we performed a three-stage blind analysis. The first second of each interval, i.e.~10\,\% of the data, was used to develop the analysis pipeline but was not included in the final spectrum. Data from seconds 2--3 of each interval were unblinded to verify that the analysis pipeline was indeed invariant under the presence of a larger statistical sample. Given that the initial unblinding satisfied this condition, we unblinded the remaining data and analyzed seconds 2--10 from each data partition, i.e.~90\,\% of the data defined as the DM-search data, to extract the final results.


To identify physics events, we triggered on pulses within the continuous-acquisition data stream offline.
To issue triggers, we first applied a matched filter to the data stream using an exponential pulse template (20\,$\mu$s rise time and 80\,$\mu$s fall time) and then
set a trigger threshold equivalent to $\sim$\,0.4~$e^{-}h^{+}$~pairs for event identification.
The event trigger time is the time at which the triggered pulse is at its maximum.
After verifying that the two QET channels on the HVeV detector have equal gain, this trigger scheme was applied to the sum of the two channels' data streams and 
separately to the veto detector.


Pulse energy and time were reconstructed using an optimal filter (OF) algorithm~\cite{Gatti:1986,Kurinsky:2018}. 
The OF algorithm requires a pulse template and the noise power spectral density (PSD). 
We constructed the pulse template for the OF algorithm from the laser-calibration event pulses with high-frequency noise filtered out. 
We measured the noise PSD on an hourly basis to account for variations of the environmental noise, using the first 100 seconds of each one-hour data partition with triggered pulses removed.
The pulse amplitude and time that minimize the frequency-domain $\chi^{2}$ were determined within a time window of [$-\,678\,\mu$s, $+\,2034\,\mu$s] centered on the trigger.
These amplitude and time quantities of the OF algorithm were also used to compute a time-domain
$\chi^{2}$ which was used later in the analysis.

Temperature fluctuations at the detector and small variations in the HV bias resulted in a small variation (${<1\,\%}$) in the detector gain.
We used the quantized $e^{-}h^{+}$-pair peaks in the aforementioned temperature-controlled sample spectrum, as well as the daily laser-calibration spectra, to linearly correct for the temperature and voltage dependencies. We then
corrected for
nonlinearities in the detector response
with a second-order polynomial.

\begin{figure}[t!]
    \centering
    \includegraphics[width=1.0\columnwidth]{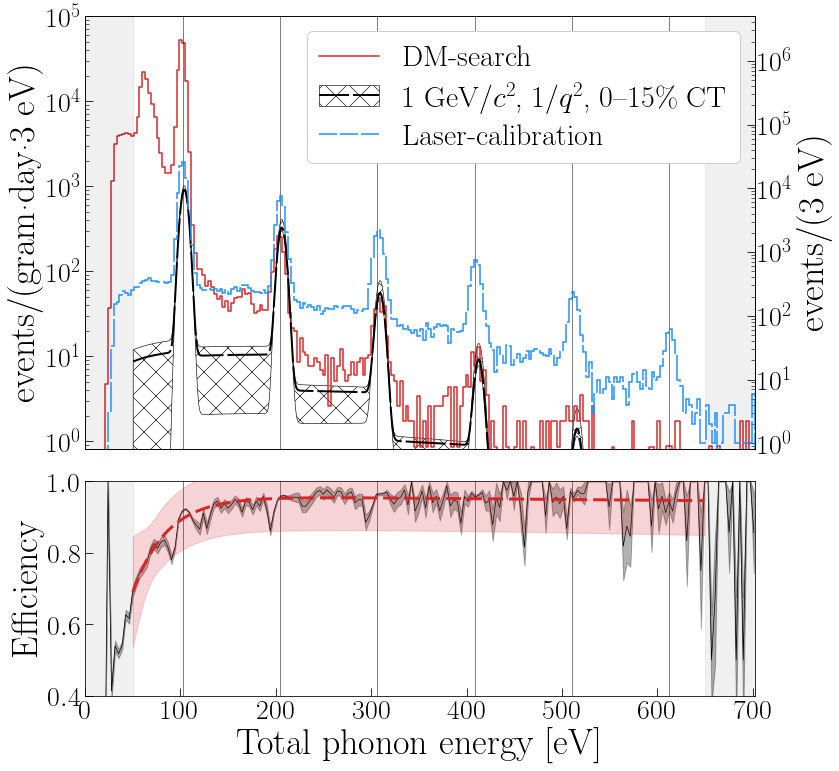}
    \caption{The top panel shows the DM-search spectrum (red) in units of event rate per 3\,eV bin (left y-axis) and the laser-calibration spectrum (blue) in units of events per 3\,eV bin (right y-axis). Both spectra show the data measured with a detector bias of 100\,V after applying the live-time and data-quality cuts. The peak seen at $\sim$\,50\,eV in the DM-search data is due to non-quantized events restricted to the outer QET channel~\cite{Hong:2019}. Light gray-shaded regions on the left- and right-hand sides mark the energy ranges outside the region of interest; vertical lines correspond to the phonon energy $E_{n}$ of the $n$-photon absorption peak (Eq.~\ref{eq:energy_scale_quantized}). The black curve is an example of a signal produced by electron-recoiling dark matter particles with a mass of 1\,GeV$/c^{2}$ and form factor $F_{\textrm{DM}} \propto 1/q^{2}$. This model assumes a Fano factor of $F=0.155$, an impact ionization (II) probability of 2\,\%, and a charge trapping (CT) probability that varies from 0--15\,\% shown by the hatched region. The curve is scaled to the dark matter-electron cross section $\bar{\sigma}_{e}$ that sets the limit at the 2$^{\textrm{nd}}$ $e^{-}h^{+}$-pair peak. The bottom panel shows the binned efficiency data $\epsilon(E_{i})$ (grey solid line), where the corresponding shaded region indicates the $1\,\sigma$ statistical uncertainty in each bin. The red dashed curve is the efficiency curve, and the corresponding shaded region is the conservative efficiency uncertainty envelope, which accounts for the statistical and systematic uncertainties.}
    \label{fig:spectrum_efficiency_100V}
\end{figure}

To calibrate the OF pulse amplitudes to energies we rescaled the $e^{-}h^{+}$-pair peaks assuming
\begin{equation}
    E_{n}= n(E_{\gamma} + e\cdot V_{\textrm{bias}}), \label{eq:energy_scale_quantized} 
\end{equation}
where $n$ denotes the $n$-photon absorption peak, ${E_{\gamma}=1.95}$\,eV is the laser photon energy, and $e$ is the absolute value of the electron charge.
Figure~\ref{fig:spectrum_efficiency_100V} (top panel) shows the resulting spectra from the DM-search and laser-calibration data.

\section{Data selection}
To ensure accurate event reconstruction,
individual live-time intervals from the DM-search and laser-calibration data were discarded (cut) based on various criteria: 
(1) the ADR temperature to exclude data outside the range of the temperature calibration; (2) the pre-pulse baseline averaged over one-second intervals to reject periods of time when the detector was still recovering from a preceding high energy deposition; and (3) the trigger rate to remove bursts of non-DM triggers. The trigger-rate cut was not applied to the laser-calibration data. The data remaining after these live-time cuts define the science exposure for this analysis, and yielded a DM-search exposure of 1.2~gram-days.

To reject poorly reconstructed events in the DM-search exposure, a set of event-by-event data-quality cuts were applied based on: (1) the difference between the OF-determined pulse time and the event trigger time to reject noise triggers and pulses affected by pile-up events; (2) the event-by-event average pre-pulse baseline to ensure the detector is at a steady working condition before an event occurs; and (3) the energy-dependent frequency- and time-domain $\chi^{2}$ quantities to remove pile-up events and baseline excursions that are unlikely to have been caused by DM-triggers. 
To define the cuts, we determined the nominal distributions of each parameter using the laser-calibration data and discarded events in the DM-search exposure exhibiting an excursion of $>3\,\sigma$ in any of these parameters. Lastly, we rejected events with a coincident triggered event in the veto detector.

The cut efficiency as a function of phonon energy was determined using the
laser-calibration
data after applying the live-time cuts to pulses coincident with the laser trigger signal. Pile-up events that occurred within a laser-event trigger window were included as part of the efficiency calculation. 
The binned efficiency $\epsilon(E_{i})$
is the fraction of events in the $i$-th bin that pass the quality cuts.
We expect the efficiency to be smooth; however, our measurement of $\epsilon(E_{i})$ shown in Fig.~\ref{fig:spectrum_efficiency_100V} (bottom panel) has both statistical fluctuations and systematic uncertainty.
In order to avoid folding these effects into the final results, we fit a smooth function to $\epsilon(E_{i})$ and assigned a conservative envelope that accounts for the statistical and systematic uncertainties. The systematic uncertainty is due to pile-up events that were not rejected by the live-time cuts, resulting in a misreconstruction of the energy. Although this envelope was propagated as part of the total experimental uncertainty in the final results, we verified that it is not the dominant source of uncertainty. 

The energy region of interest (ROI) for this analysis is 50--650\,eV.
The lower bound guarantees inclusion of the full single-$e^{-}h^{+}$-pair peak at 100~V bias and a trigger efficiency consistent with
unity.
We set the upper bound at 650\,eV to focus on the corresponding low-mass ranges where this analysis has competitive sensitivity.



\section{Results} \label{sec:results}
\subsection{DM Signal Models}
The blinded DM-search data were analyzed to set exclusion limits on light DM $\chi$ scattering as well as dark photon $V$ and axion-like particle (ALP) absorption. The DM models for $\chi$, $V$, and ALPs are identical to those used in Ref.~\cite{Agnese:2018} and~\cite{Aralis:2019}. We set limits on the kinetic mixing parameter $\varepsilon$ for $V$, the axio-electric coupling $g_{ae}$ for ALPs, and the effective DM-electron cross section $\bar{\sigma}_{e}$ for $\chi$. In all cases we assume that the respective DM candidate constitutes all of the galactic DM with a local mass density of $\rho_{\textrm{DM}} = 0.3$~GeV$c^{-2}$cm$^{-3}$.

The $V$ and ALP absorption rates are proportional to the photoelectric absorption cross section $\sigma_{pe}$ of the Si detector. Discrepancies in the literature for $\sigma_{pe}$ \cite{Hildebrandt:1973,Brown:1972,Henke:1993,DaChun:1995,Green:1995,Green:2008,Dash:1955,Edwards:1997,Macfarlane:1958,Holland:2003,Gullikson:1994,XCOM,Brown:1977,Hunter:1966,Aspnes:1983,Hulthen:1975} for regions within our analysis range led us to define nominal, upper, and lower photoelectric cross-section curves to accommodate the range of published values. 
The nominal $\sigma_{pe}$ curve follows the approach taken in Ref.~\cite{Hochberg:2017}, with data from Ref.~\cite{Edwards:1997} for photon energies below 1\,keV. 

The upper and lower $\sigma_{pe}$ curves are derived from tracing upper and lower bounds of the published data after applying temperature corrections, along with the nominal curve data that did not have temperature corrections applied. The corrections account for the temperature dependence of indirect, phonon-assisted absorption that occurs at energies below the direct band gap ($\sim$3\,eV). We followed the methodology and analytical model for photon absorption found in Ref.~\cite{Rajkanan:1979} to extrapolate the data below 4\,eV to a temperature of 50\,mK.

This analysis adopted the same ionization production model as used in Ref.~\cite{Agnese:2018} to compute the mean number of $e^{-}h^{+}$ pairs $n_{eh}$ produced for an interaction with a given recoil/absorption energy. 
For recoil/absorption energies above the Si band gap $E_{gap} = 1.2$\,eV but below the average energy per $e^{-}h^{+}$ pair $\epsilon_{eh} = 3.8$\,eV, $n_{eh}=1$;
for energies above $\epsilon_{eh}$, we determined $e^{-}h^{+}$ pair probabilities from binomial distributions using selected Fano factor values, $F$. 


The total phonon energy measured for an event, $E_{ph}$, is the recoil/absorption energy $E_{r}$ plus the energy produced by the NTL effect:
\begin{equation}
    E_{ph} = E_{r} + n_{eh} \cdot e \cdot V_{\textrm{bias}}
\end{equation}
where the ionization production model and Fano statistics determine the distribution for $n_{eh}$. We combined the signal models with a charge trapping (CT) and impact ionization (II) likelihood model, which mainly contributes to the distribution of events between quantized $e^{-}h^{+}$-pair peaks~\cite{Ponce:2020}. Charge trapping occurs when an electron or hole falls into a charge vacancy in the crystal, reducing the total number of electrons or holes that traverse the entire detector and lowering the measured energy for an event. Impact ionization occurs when a moving charge in the crystal liberates an additional loosely bound charge, thereby increasing the measured energy for an event. 

We determined the CT and II probabilities by fitting the model used in Ref~\cite{Ponce:2020} to the laser-calibration data. The results from the fit are $11 \pm 3\,\%$ and $2^{+3}_{-2}\,\%$ for the CT and II probabilities, respectively, and were subsequently used to generate the signal models. 
Because we were unable to determine an energy dependence of the energy resolution within the ROI for this analysis (50--650~eV), the signal models were convolved with a weighted average energy resolution: $\sigma_{\left< E \right>} = 3.6$\,eV. We determined $\sigma_{\left< E \right>}$ by averaging over the resolutions of the first six, Gaussian-fitted $e^{-}h^{+}$-pair peaks from the combined laser-calibration data weighted by the corresponding uncertainty in each peak.
    Lastly, we multiplied each signal model by the efficiency curve (bottom panel of Fig.~\ref{fig:spectrum_efficiency_100V}) as well as the exposure (1.2~gram-days). An example of a 1\,GeV$/c^{2}$ light DM signal model is shown in the top panel of Fig.~\ref{fig:spectrum_efficiency_100V}.

\begin{figure}[htb!] 
    \centering
    \includegraphics[width=1.0\columnwidth]{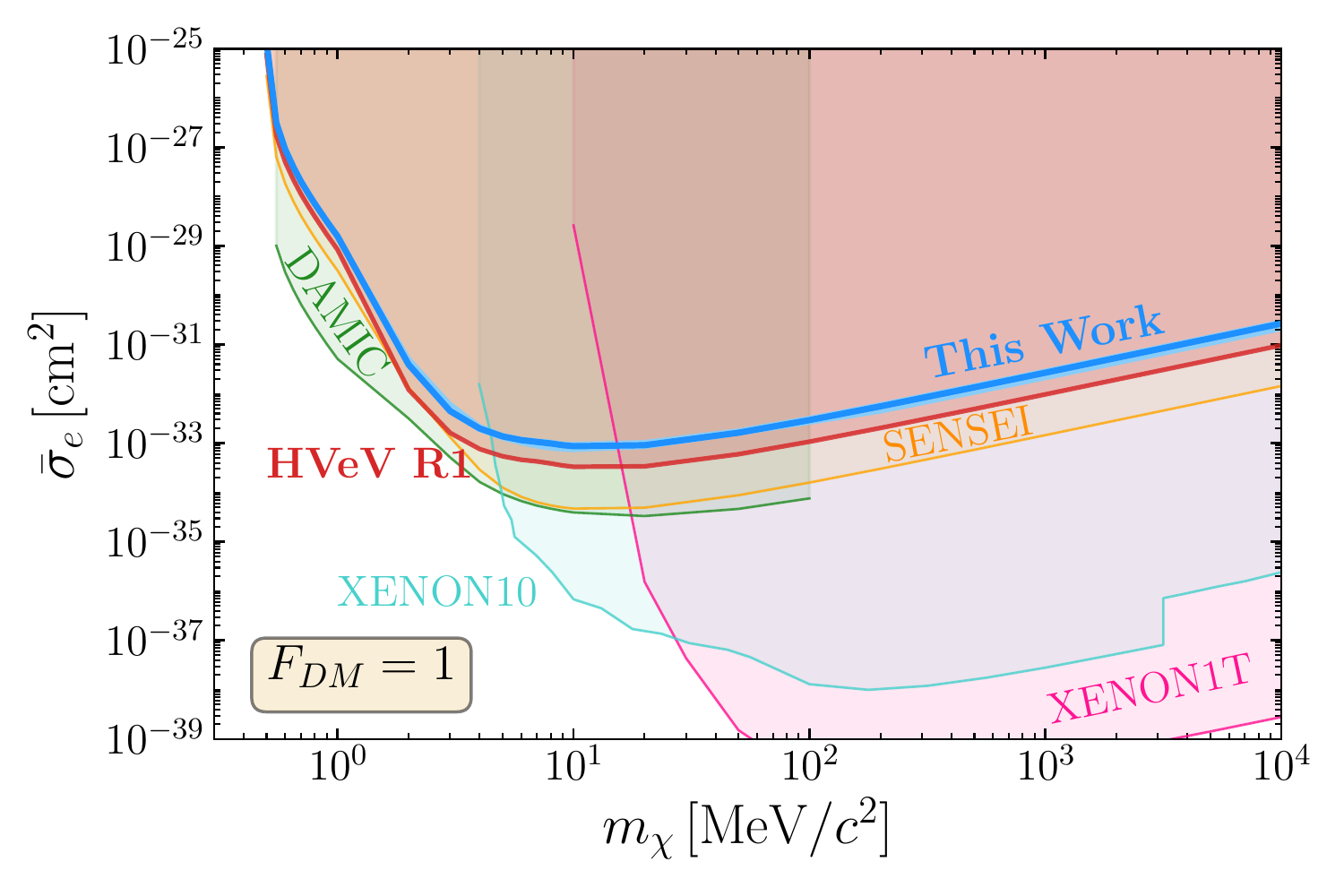}\\
    \includegraphics[width=1.0\columnwidth]{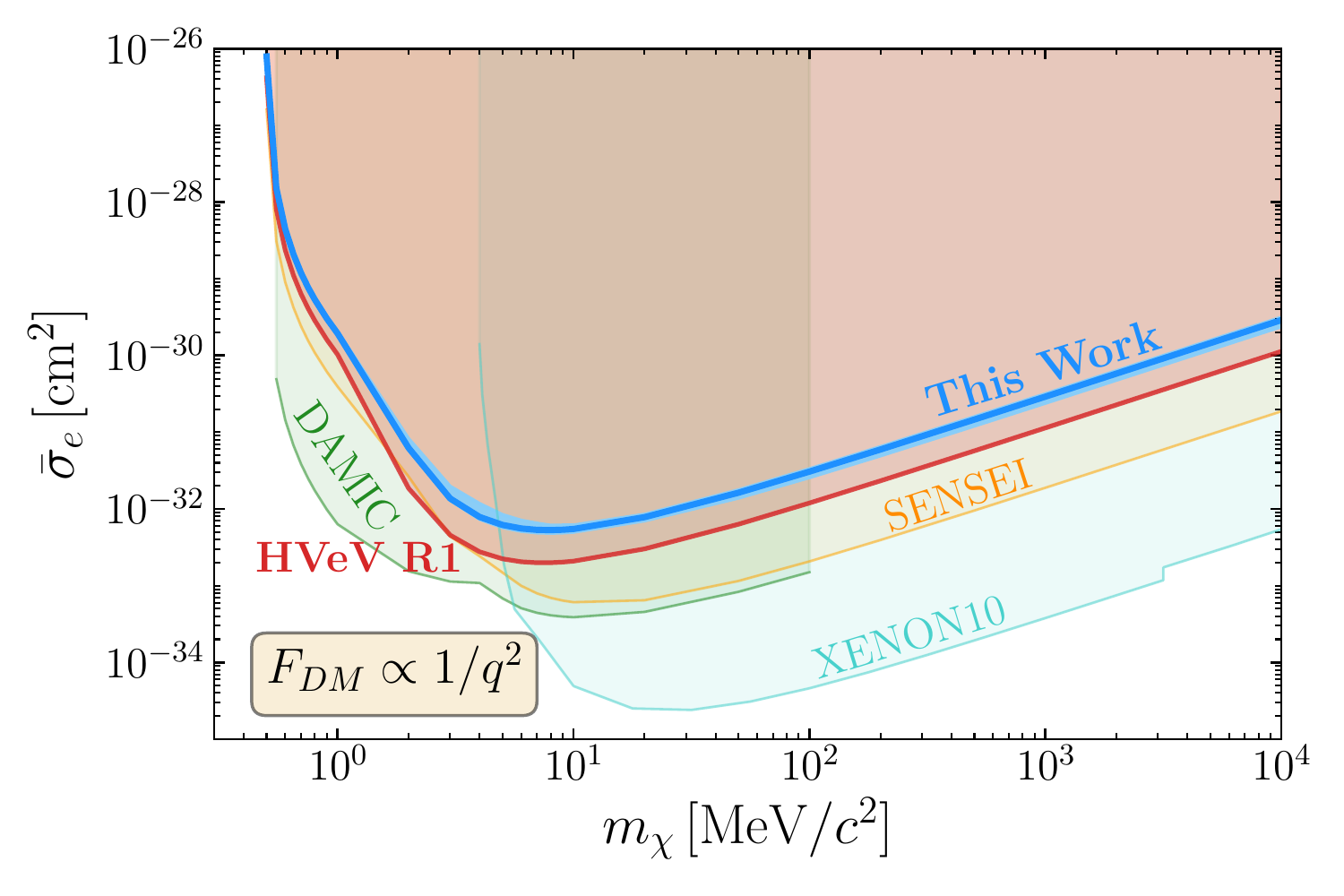}\\
    \caption{90\,\% C.L. limits on the effective dark matter-electron scattering cross section with form factor $F_{\textrm{DM}}=1$ (top) and $F_{\textrm{DM}}\propto 1/q^{2}$ (bottom) and with Fano factor of 0.155 (solid-blue curve). The light blue band represents our estimate of the systematic uncertainty, which is dominated by varying the Fano factor assumption in the ionization model from $F$ = 10$^{-4}$ to 0.3. Other direct detection constraints shown include SuperCDMS HVeV~R1~\cite{Agnese:2018} (red), DAMIC~\cite{Aguilar-Arevalo:2019} (green), SENSEI~\cite{Abramoff:2019} (orange), XENON10~\cite{Essig:2017,Essig:2012} (teal), and XENON1T~ \cite{Aprile:2019} (pink).}
    \label{fig:limit_dme}
\end{figure}

\begin{figure}[htb!]
    \centering
    \includegraphics[width=1.0\columnwidth]{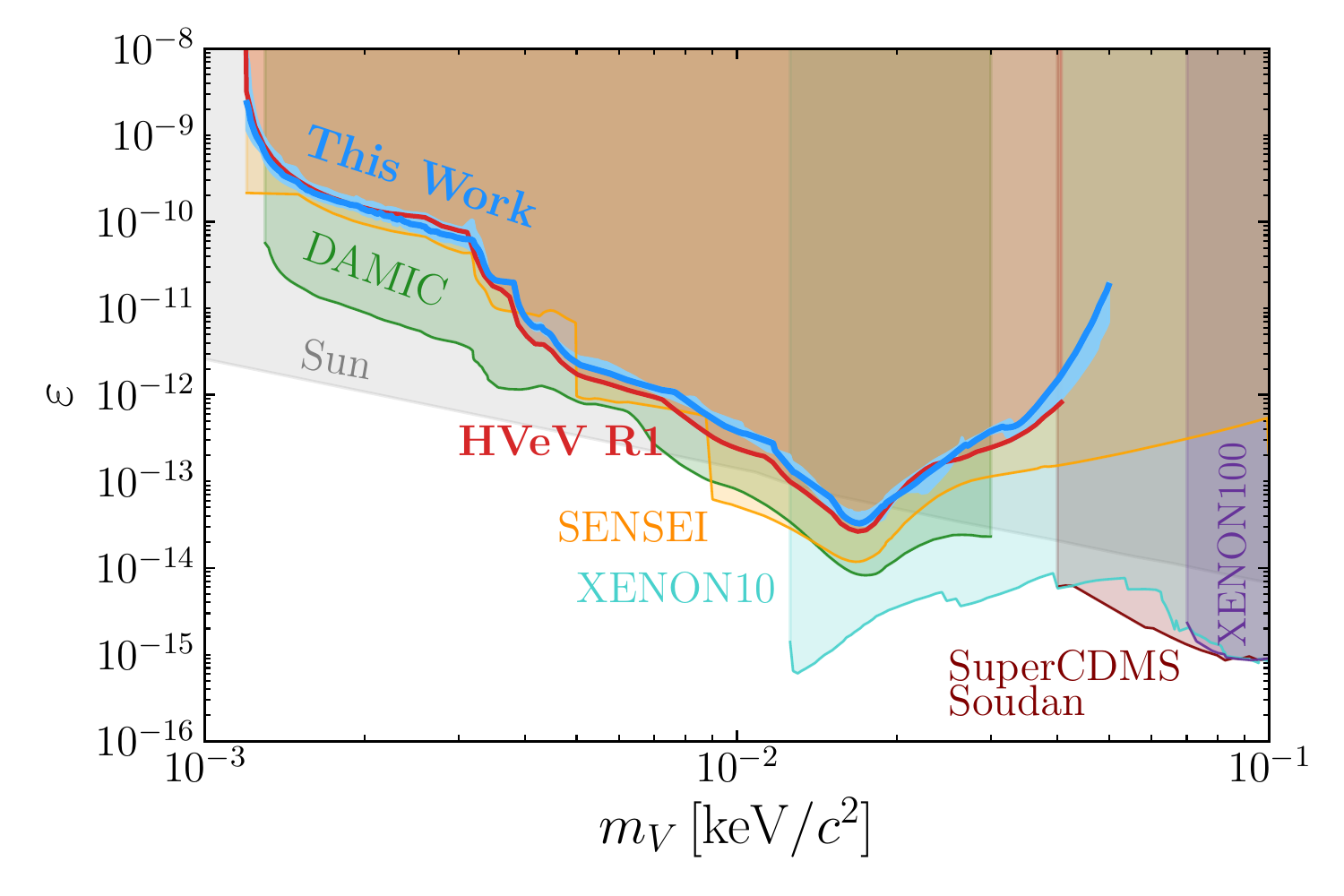}\\
    \includegraphics[width=1.0\columnwidth]{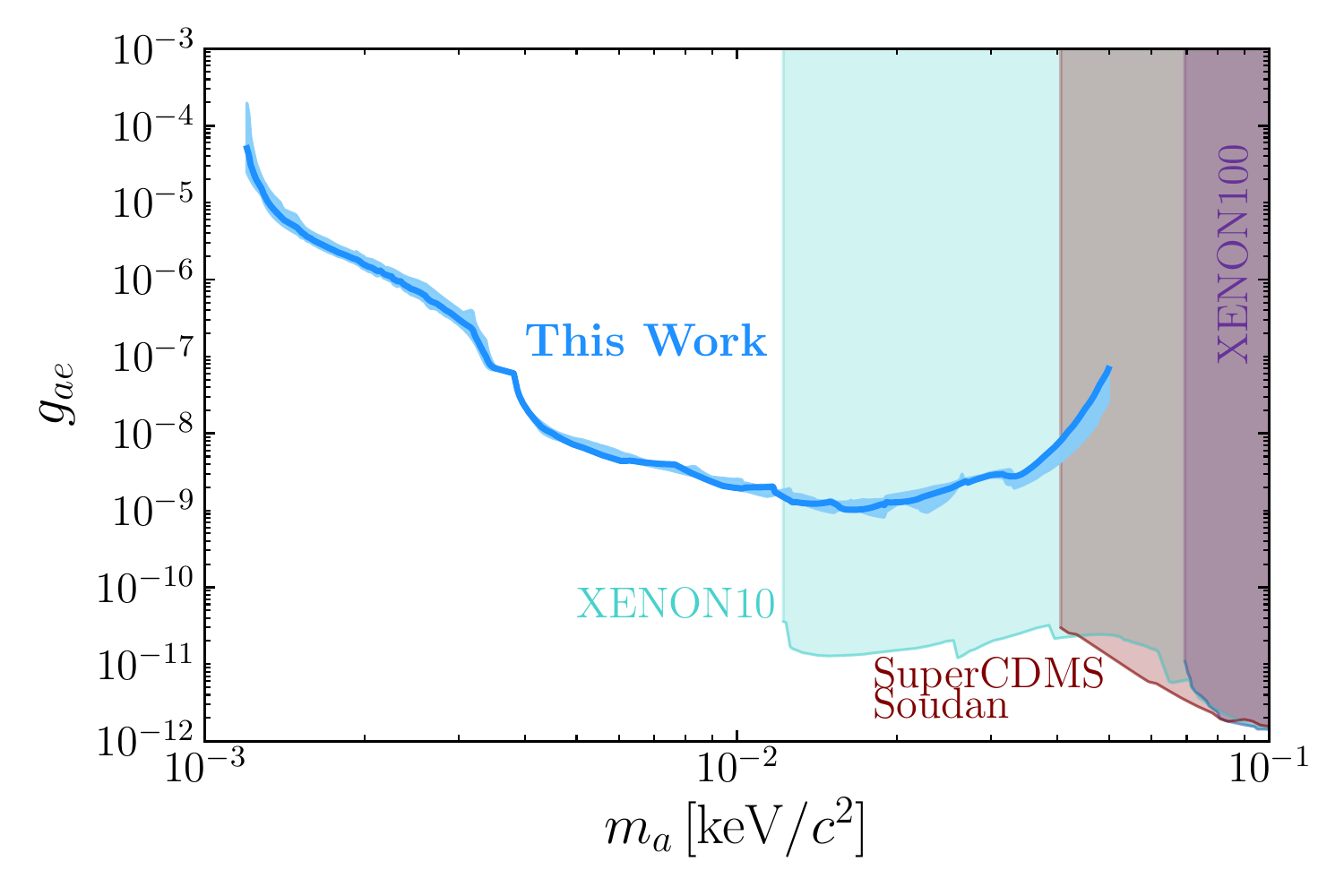}\\
    \caption{90\,\% C.L. limits on the dark photon ($V$) kinetic mixing parameter $\varepsilon$ (top) and axioelectric coupling constant $g_{ae}$ (bottom) with Fano factor of 0.155 (solid-blue curve). The light blue band represents our estimate of the systematic uncertainty, which for masses $\gtrsim$ $4\times10^{-3}$\,keV$/c^{2}$ is dominated by varying the Fano factor assumption in the ionization model from $F$\,=\,10$^{-4}$ to 0.3; for masses $\lesssim$ $4\times10^{-3}$\,keV$/c^{2}$, the uncertainty is dominated by the discrepancy in the photoelectric absorption cross section. Other direct detection constraints shown for $V$ and ALPs include SuperCDMS Soudan~\cite{Aralis:2019} (maroon), XENON10 (teal), and XENON100 (purple)~\cite{Bloch:2017}; additional constraints on $V$ include SuperCDMS HVeV~R1~\cite{Agnese:2018} (red), DAMIC~\cite{Aguilar-Arevalo:2019} (green), SENSEI~\cite{Abramoff:2019} (orange), and anomalous energy loss mechanisms in the Sun~\cite{An:2015}. For the axioelectric coupling, the entire region shown is disfavored by the observed cooling of red giant~\cite{Viaux:2013, Tanabashi:2018} and white dwarf stars~\cite{Bertolami:2014,Tanabashi:2018}.}
    \label{fig:limit_abs}
\end{figure}

\subsection{Limit Setting}
The Poisson exclusion limit for each DM model was calculated independently for the first six $e^{-}h^{+}$-pair peaks using a limit setting window of $\pm~3\,\sigma_{\left< E \right>}$ centered on each peak. While taking into account the look-elsewhere effect, we selected the lowest limit amongst the individual $e^{-}h^{+}$-pair peaks at each DM mass to obtain a final limit with a  90\,\% confidence level (C.L.). 

This limit calculation differs from Ref.~\cite{Agnese:2018}, which determined the limits using the Optimum Interval (OI) method \cite{Yellin:2002,Yellin:2007}. 
Due to the improved energy resolution of this analysis compared to Ref.~\cite{Agnese:2018}, the OI method was found to be overly sensitive to the shape of the expected DM signals measured in the detector and thus to the effects of CT and II, leading to systematic uncertainties that are difficult to estimate. In contrast, the Poisson method applied to this analysis is insensitive to these systematic effects.
A comparison of the two methods finds up to a factor of 2 stronger limits with the OI method due to the sensitivity to the model shape (the same comparison performed on the Ref.~\cite{Agnese:2018} analysis results in no such difference due to the poorer energy resolution). 
In this analysis we used the more conservative Poisson limit setting method, as it is more effective at constraining the systematic uncertainties.


To quantify the impact of systematic uncertainties, the limits were recalculated with Gaussian distributed random variates for the energy calibration, energy resolution, CT and II fractions, and efficiency, according to their corresponding means and uncertainties. For the photoelectric absorption cross section, we made a random choice between the lower, upper, and nominal curves. At each mass, we took the average from all trials and used the $\pm\,1\,\sigma$ equivalent values from the resulting limit distribution as the limit uncertainty. The limits and their propagated uncertainty are calculated separately using three different values for the Fano factor: the one measured at high energy, $F=0.155$ \cite{Owens:2002}, and the values of $F=10^{-4}$ and $F=0.3$ assumed to cover the systematic uncertainty of the Fano factor at these energies.

Figures~\ref{fig:limit_dme} and~\ref{fig:limit_abs} show the limits on $\chi$ scattering and $V$/ALP absorption, respectively, compared to existing limits. The limits on $\chi$ assume a DM form factor of either $F_{\textrm{DM}} = 1$ or $F_{\textrm{DM}} \propto 1/q^{2}$~\cite{Essig:2016}. The light blue bands representing our estimates of the systematic uncertainty envelops the $\pm1\,\sigma$ values of all three limits obtained using the different Fano factor assumptions in the ionization model. At most masses, the uncertainty bands are dominated by the varying Fano factor assumption;
the exception is for $\lesssim 4$\,eV$/c^{2}$ in the $V$ and ALP absorption models, where the uncertainty is dominated by the discrepancy in the photoelectric absorption cross section.


\section{Discussion and Outlook} \label{sec:disout}
The limits in Figs.~\ref{fig:limit_dme} and~\ref{fig:limit_abs} are remarkably close to those from our previous run~\cite{Agnese:2018} despite the $\sim$\,2.5 times larger exposure. 
They are in fact weaker at some masses due to the higher observed event rate in the 3$^{\textrm{rd}}$ $e^{-}h^{+}$-pair peak coupled with the higher CT and II probabilities in this measurement, as well as the use of a more conservative limit setting method (see Ref.~\cite{Ponce:2020} for recent CT and II measurements of the detector used in Ref.~\cite{Agnese:2018}). 
Table~\ref{tab:eventrate_compare} compares the efficiency-corrected event rates for each $e^{-}h^{+}$-pair peak within a $\pm~3\,\sigma_{E}$ window. The event rate observed in each peak is similar in this run compared to Ref.~\cite{Agnese:2018} despite a different detector design, cryostat, location, overburden, and shielding.

Another 0.39~gram-days of data were taken with a bias of 60\,V across the detector in order to determine if the results are voltage-dependent. Table~\ref{tab:eventrate_compare} shows that the resulting event rate for each number of $e^{-}h^{+}$ pairs is similar to the corresponding rate from the 100\,V data, suggesting a voltage-independent result.
Furthermore, the event rate above the first $e^{-}h^{+}$-pair peak is comparable to that seen in other charge-readout experiments~\cite{Agnese:2018,Abramoff:2019,Aguilar-Arevalo:2019,collaboration2020germaniumbased}, and adds to the growing narrative of unexplained, $\mathcal{O}(\textrm{Hz/kg})$ low-energy excesses measured in many sub-GeV DM searches (Refs.~\cite{Kurinsky:2020,robinson2020comment,kurinsky2020reply} and references therein).
This result from our detector with unparalleled energy resolution provides a new dataset that can contribute to understanding the origin of these unknown background events. 
A third run with an identically designed detector is planned in a dilution refrigerator in a shallow underground site with 255~m.w.e.\ overburden (NEXUS Facility~\cite{Battaglieri:2017}) to probe the correlation between the unknown events and known environmental background sources.

\begin{table}[tb]
\caption{Comparison of the efficiency-corrected event rate in each $e^{-}h^{+}$-pair peak between this work and Ref.~\cite{Agnese:2018}. The event rates displayed from this analysis are calculated from the DM-search data measured with a bias voltage of 100\,V, as well as from the additional dataset measured with a bias voltage of 60\,V. For each number of $e^{-}h^{+}$ pairs, the event rate is determined by counting the number of observed events within a $\pm~3\,\sigma_{E}$ window centered on the peak. The uncertainty shown is the 3\,$\sigma$ uncertainty in the number of observed events assuming Poisson statistics.}
\begin{tabular}{rccc}
\toprule
\multicolumn{1}{l}{} & \multicolumn{2}{c}{This Work}               & Ref.~\cite{Agnese:2018}             \\
\cline{2-3} \cline{4-4}
Voltage [V]          & 100 & 60 & $-140$     \\ 
$\sigma_{E}$ [$e^{-}h^{+}$]          & 0.03  & 0.05 & 0.1 \\ \hline
     &   \multicolumn{3}{c}{Events/(gram-day)}                 \\ 
\cline{2-4}
1 $e^{-}h^{+}$         & $(149\pm 1)\,10^3$             & $(165\pm 2)\,10^3$            & $(157 \pm 2)\,10^3$                    \\ 
2 $e^{-}h^{+}$         & $(1.1 \pm 0.1)\,10^3$             & $(1.2 \pm 0.2)\, 10^3$            & $(1.3 \pm 0.2)\,10^3$                    \\ 
3 $e^{-}h^{+}$         & $207 \pm 40$             & $245 \pm 86$            & $171 \pm 59$                    \\ 
4 $e^{-}h^{+}$         & $53 \pm 20$             & $77 \pm 48$            & $58 \pm 34$                    \\ 
5 $e^{-}h^{+}$         & $16 \pm 11$             & $20 \pm 25$            & $16 \pm 18$                \\ 
6 $e^{-}h^{+}$         & $5 \pm 6$         & $10 \pm 17$        & $24 \pm 22$                    \\ \botrule
\end{tabular}

\label{tab:eventrate_compare}
\end{table}

Finally, due to the significant impact that charge trapping and impact ionization have on the signal reconstruction, there is an ongoing effort toward understanding these charge propagation effects and investigating factors that influence them. A DM model spectrum with CT and II included is shown in the top panel of Fig~\ref{fig:spectrum_efficiency_100V}. The black curve shows the DM signal model for a 1\,GeV$/c^{2}$ light DM particle with form factor $F_{\textrm{DM}} \propto 1/q^{2}$, scaled to the limit of excluded $\bar{\sigma}_{e}$, using an II probability of 2\,\%. The CT probability shown by the hatched region is varied from 0--15\,\%. For this model and limit-setting scheme, these processes do not determine the ultimate sensitivity. However, lower CT and II rates combined with a more robust understanding~\cite{MoffattAPL:2019,StanfordAIP:2020} will allow us to use the region between the peaks in the limit-setting procedure to improve the sensitivity of future analyses, as well as to fully utilize the improved resolution of this detector for additional background discrimination.

\section{Acknowledgements}

We would like to thank Rouven Essig and Tien-Tien Yu for helpful discussions and assistance with using QEdark \cite{Essig:2016} to generate the dark matter model used in this analysis. We thank
Noemie Bastidon for her work in the preliminary design of our optical fiber setup and wire bonding.
We gratefully acknowledge support from the U.S. Department of Energy (DOE) Office of High Energy Physics and from the National Science Foundation (NSF).  This work was supported in part under NSF Grants No.\,1809730 and No.\,1707704, as well as by the Arthur B. McDonald Canadian Astroparticle Physics Research Institute, NSERC Canada, the Canada Excellence Research Chair Fund, Deutsche Forschungsgemeinschaft (DFG, German Research Foundation) under Project No.\,420484612 and under Germany’s Excellence Strategy - EXC 2121 ``Quantum Universe" - 390833306, and the Department of Atomic Energy Government of India (DAE) under the project - Research in basic sciences (Dark matter). Fermilab is operated by Fermi Research Alliance, LLC, under Contract No.\,DE-AC02-37407CH11359 with the US Department of Energy. Pacific Northwest National Laboratory (PNNL) is operated by Battelle Memorial Institute for the DOE under Contract No.\,DE-AC05-76RL01830. SLAC is operated under Contract No.\,DEAC02-76SF00515 with the United States Department of Energy. 

\bibliography{refs}

\end{document}

%% file: authors.tex
\author{D.W.~Amaral} \affiliation{Department of Physics, Durham University, Durham DH1 3LE, UK}
\author{T.~Aralis} \affiliation{Division of Physics, Mathematics, \& Astronomy, California Institute of Technology, Pasadena, CA 91125, USA}
\author{T.~Aramaki} \affiliation{SLAC National Accelerator Laboratory/Kavli Institute for Particle Astrophysics and Cosmology, Menlo Park, CA 94025, USA}
\author{I.J.~Arnquist} \affiliation{Pacific Northwest National Laboratory, Richland, WA 99352, USA}
\author{E.~Azadbakht} \affiliation{Department of Physics and Astronomy, and the Mitchell Institute for Fundamental Physics and Astronomy, Texas A\&M University, College Station, TX 77843, USA}
\author{S.~Banik} \affiliation{School of Physical Sciences, National Institute of Science Education and Research, HBNI, Jatni - 752050, India}
\author{D.~Barker} \affiliation{School of Physics \& Astronomy, University of Minnesota, Minneapolis, MN 55455, USA}
\author{C.~Bathurst} \affiliation{Department of Physics, University of Florida, Gainesville, FL 32611, USA}
\author{D.A.~Bauer} \affiliation{Fermi National Accelerator Laboratory, Batavia, IL 60510, USA}
\author{L.V.S.~Bezerra} \affiliation{Department of Physics \& Astronomy, University of British Columbia, Vancouver, BC V6T 1Z1, Canada}\affiliation{TRIUMF, Vancouver, BC V6T 2A3, Canada}
% ...
